\documentclass[10pt,conference]{IEEEtran}
\IEEEoverridecommandlockouts
\usepackage{amsmath,amssymb,amsfonts, amsthm}
\usepackage{algorithmic}
\usepackage{graphicx}
\usepackage{textcomp}
\usepackage{xcolor}
\usepackage{makecell}
\usepackage{booktabs}
\usepackage{subfigure}
\usepackage{url}
\usepackage{flushend}
\DeclareUnicodeCharacter{0301}{\'{e}}
\usepackage[backend=biber, sorting=none, style=ieee, url=false]{biblatex}
\def\BibTeX{{\rm B\kern-.05em{\sc i\kern-.025em b}\kern-.08em
    T\kern-.1667em\lower.7ex\hbox{E}\kern-.125emX}}

\theoremstyle{definition}
\newtheorem{definition}{Definition}[section]
\begin{document}

\title{Data-Flow-Based Normalization Generation Algorithm of R1CS for Zero-Knowledge Proof}

\author{\IEEEauthorblockN{Chenhao Shi, Hao Chen, Ruibang Liu, Guoqiang Li\dag}\thanks{\dag\,Corresponding author.}
\IEEEauthorblockA{\textit{Shanghai Jiao Tong University, Shanghai 200240, China}\\
\{undefeated, Dennis.Chen, 628628, li.g\}@sjtu.edu.cn}
}

\maketitle

\begin{abstract}
The communities of blockchains and distributed ledgers have been stirred up by the introduction of zero-knowledge proofs (ZKPs). Originally designed to solve privacy issues, ZKPs have now evolved into an effective remedy for scalability concerns and are applied in Zcash (internet money like Bitcoin). To enable ZKPs, Rank-1 Constraint Systems (R1CS) offer a verifier for bi-linear equations. To accurately and efficiently represent R1CS, several language tools like Circom, Noir, and Snarky have been proposed to automate the compilation of advanced programs into R1CS. However, due to the flexible nature of R1CS representation, there can be significant differences in the compiled R1CS forms generated from circuit language programs with the same underlying semantics. To address this issue, this paper uses a data-flow-based R1CS paradigm algorithm, which produces a standardized format for different R1CS instances with identical semantics. By using the normalized R1CS format circuits, the complexity of circuits' verification can be reduced. In addition, this paper presents an R1CS normalization algorithm benchmark, and our experimental evaluation demonstrates the effectiveness and correctness of our methods.
\end{abstract}


\section{Introduction}
\emph{Zero-knowledge proofs (ZKPs)} are increasingly recognized for their importance in modern cryptography~\cite{goldwasser2019knowledge}, as one and more cryptographic communities seek to address some of the blockchain's most significant challenges: privacy and scalability. It is also the essential technique in Zcash~\cite{zcash, sasson2014zerocash}. From both user's and developer's perspectives, the heightened emphasis on information privacy and security has led to a greater appreciation for the privacy advantages offered by zero-knowledge proofs. As decentralized finance (DeFi) usage grows, zero-knowledge applications that provide scalability and privacy advantages will have more opportunities to increase industry-wide adoption. However, not all computational problems can be directly addressed using zero-knowledge proofs. Instead, we must transform the issue into the correct form of computation. The \emph{rank-1 constraint system (R1CS)} describes the execution of statements written in high-level programming languages and is used by many ZKP applications, but there is no standard way of representing them~\cite{dre2019jr1cs}.  Circom is a novel domain-specific language for transforming computational problems into R1CS format circuits.\cite{albert2022distilling} In the specific process of a first-order zero-knowledge proof, we first convert the problem into a computational problem in Circom, then into R1CS format circuits.


Due to the flexible nature of R1CS representation and variation of program organizations and compiler optimization levels, there can be significant differences in the compiled R1CS forms generated from circuit language programs with the same underlying semantics, which leads to difficulties in further ZKP program analysis and verification. 



This paper proposes a data-flow-based algorithm for generating normalization of R1CS, enabling the conversion of different R1CS constraints into a normal form, facilitating the determination of equivalence and correctness. To achieve this, the algorithm starts by transforming an R1CS into a data flow graph structure resembling an expression tree. It then segments and abstracts the data flow graph, eliminating differences between equivalent R1CS constraints that may arise from the generation process. Finally, sorting rules are proposed to sort the constraints and variables within R1CS, ultimately resulting in a unique normal form for equivalent R1CS.


Moreover, we classify and summarize the reasons and characteristics of the different equivalent R1CS generated, based on the constraint generation logic of mainstream compilers and the expressiveness of R1CS. In addition, based on the identified reasons for producing equivalent R1CS, we create a relatively complete benchmark. Our proposed algorithm, which can pass all test cases in the benchmark, demonstrates that equivalent R1CS can be converted into a unique and identical canonical form under various circumstances.


This work contributes to R1CS optimization by providing a novel algorithm for generating canonical forms of equivalent R1CS constraints. Our algorithm can eliminate unnecessary redundancy and normalize representation, thus it can improve existing methods and facilitates the analysis of equivalence and correctness. Furthermore, the effectiveness and practicality of the proposed algorithm are demonstrated through our comprehensive benchmark.


\noindent
\textbf{Related Work}
Eli et al. design, implement, and evaluate a zero-knowledge succinct non-interactive argument (SNARG) R1CS~\cite{ben2019aurora}. Historically, research investigating the factors associated with R1CS has focused on satisfiability. In paper\cite{lee2021linear}, where the prover incurs finite field operations to prove the satisfiability of an n-sized R1CS instance~\cite{lee2021linear}. Alexander et al. introduce Brakedown, the first built system that provides linear-time SNARKs for NP~\cite{golovnev2021brakedown}. 

Considering Circom and R1CS format circuits as two languages before and after compilation, research on the generation of the R1CS paradigm is more akin to research on semantic consistency in the compilation. Currently, the patent applications and research papers propose ideas and solutions for generating compilation paradigms in other languages, mainly exploring data flow~\cite{yuan2019Semantic}, syntax tree~\cite{zhao0Compiling}, or semantic mapping~\cite{gao0A} aspects. These studies offer crucial insights into the fundamental information semantically identical programs entail in the compilation process. However, due to the inherent constraints embedded within the R1CS form, this paper ultimately elects to use data flow as a starting point for research.

\noindent
\textbf{Paper Organization}
The paper is organized as follows: In the next section, a brief preliminary review of zero-knowledge proof and related tools. Section 3 provides the process of the proposed algorithm in this paper. The technical exposition in Section 4 explains in detail the logic of the critical steps and their formal description. Section 5 presents the specific categories of benchmarks and their corresponding experimental results. Lastly, Section 6 concludes the present study.

\section{Preliminaries}
This section introduces the basic concepts and principles of zero-knowledge proof and discusses the role of R1CS and Circom in zero-knowledge-proof systems. It also explores the limitations of existing normalization techniques for R1CS and presents the proposed data-flow-based normalization generation algorithm, which is motivated by these limitations.

\subsection{ZK-SNARKS}
Zk-SNARK, which stands for Zero-Knowledge Succinct Non-interactive Argument of Knowledge, is a type of zero-knowledge proof introduced in a 2014 paper~\cite{ben2014succinct}. The objective of zk-SNARK is to enable one party to prove to another that they possess specific knowledge without revealing the knowledge itself, and to do so concisely and efficiently. The working principle of Zk-SNARK can be simplified into several steps. First, users convert the information they wish to verify into a mathematical problem called a computation. This computation can be implemented using any high-level programming language, such as OCaml, C++, Rust, or hardware description languages like Circom.

\begin{figure}[!ht]
    \begin{center}
    \includegraphics[width = .8\linewidth]{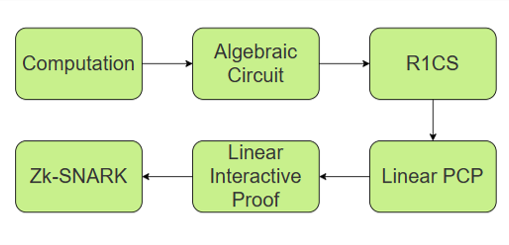}
    \caption{The pipeline diagram for zk-SNARKS.}
    \label{fig:zkp_pipeline}
    \end{center}
\end{figure}

Next, the computation result is usually transformed into an arithmetic circuit, a computational model for computing polynomials. An arithmetic circuit consists of inputs and multiple gates, performing an essential arithmetic operation such as addition or multiplication. The entire arithmetic circuit can be used to generate a specific format (R1CS), which takes the form of a constraint-based formula system (rank-1 constraint system) expressing the constraints of the arithmetic circuit. A verifiable arithmetic circuit is one of the inputs of the zk-SNARKs algorithm.

The Quadratic Arithmetic Program (QAP) is a variant of linear pcp and plays a crucial role in the verifiable arithmetic circuit's conversion into the QAP format. QAP is a formula system that employs polynomials to represent the behavior of the arithmetic circuit. By utilizing QAP, the security of the zk-SNARKs algorithm is enhanced, along with the improvement of its implementation efficiency \cite{buterin2016quadratic}.

Finally, there is the zk-SNARK stage, where the verifiable arithmetic circuit and QAP are used to generate a proof. Zk-SNARKs are powerful privacy protection protocols that can be utilized in digital payments, blockchain technology, and other fields. They can verify the authenticity of information while protecting the user's privacy. Despite its relatively complex working principle, zk-SNARK technology has found widespread application, bringing higher security and privacy protection to the digital world.

\subsection{Circuit Language}
ZKP technology can address several fundamental issues in the modern digital world, including identity user's verification without compromising their private information and safeguarding privacy data from unauthorized exploitation. Within ZKP, arithmetic circuits play a vital role in describing and computing complex operations. These circuits consist of a series of logic gates that can perform various arithmetic operations, such as addition, multiplication, and division. By combining these basic operations, the complex arithmetic circuits can be constructed to execute diverse computational tasks.

The circuit referred to here is a theoretical computational model, not an actual electronic circuit.

This is the formal arithmetic circuit definition in theoretical computer science \cite{vollmer1999introduction}.
\begin{definition} \label{def:finite_field}
    A finite field field \(\mathbb{F}\) is a field that contains a finite number of integer elements.
    \[\mathbb{F} = \{0, \ldots, p - 1\} \text{ for some prime } p > 2 \]
    The operations \(+,\times, =\) on \(\mathbb{F}\) should \(\pmod p\) after calculation.
\end{definition}

\begin{definition}\label{def:circuit_def}
A circuit is a triple $(M,L,G)$, where
\begin{enumerate}
    \item[$\cdot$] $M$ is a set of values,
    \item[$\cdot$] $L$ is a set of gate labels, each of which is a function from $M^{i}$ to $M$ for some non-negative integer $i$ (where $i$ represents the number of inputs to the gate), and
    \item[$\cdot$] $G$ is a labelled directed acyclic graph with labels from $L$.
\end{enumerate}
\end{definition}

\begin{definition} \label{def:arithmetic_circuit_def}
    A arithmetic circuit is a map \(C: \mathbb{F}^n \rightarrow \mathbb{F}\), where \(\mathbb{F}\) is a finite field.

    \begin{enumerate}
        \item It is a directed acyclic graph (DAG) where internal nodes are labeled \(+,-, \text{or } \times\) and inputs are labeled \( 1, x_1, \ldots, x_n \), the edges are wires or connections.
        \item It defines an n-variable polynomial with an evaluation recipe.
        \item Where \(|C| = \# \text{ gates in C}\).
    \end{enumerate}
\end{definition}

This chapter mainly focuses on the Circom language, employed at the core step of zk-SNARK protocols for describing arithmetic circuits. Within the framework of ZKP systems, several commonly used arithmetic circuit description languages exist, including Arithmetica, libsnark DSL, and Circom. These languages are typically employed for building and verifying ZKP systems and can be used to describe various arithmetic circuits, such as linear constraint systems (LCS), bilinear pairings, and quadratic circuits.

By representing an arithmetic circuit as a constraint system, Circom's core idea involves describing inputs, outputs, and computation processes as linear equations and inequalities. This approach allows developers to define complex computation processes and generate the corresponding R1CS constraint system. Circom provides a set of high-level abstract concepts, enabling developers to focus more on the algorithm without being overwhelmed by low-level implementation details.

Due to its comprehensive range of features, Circom has been widely used in cryptography, blockchain, and other security-critical applications. Its efficient and streamlined development of complex computational structures gives developers a chance to quickly implement various privacy-preserving protocols and zero-knowledge-proof techniques. Circom provides a simple, declarative way of defining constraint systems and describing real-world puzzles, which is of great importance to the protection of user's privacy and data security.

\subsection{Rank-1 Constraint Systems (R1CS)}

R1CS, a common Arithmetic Circuit format that underlies real-world systems~\cite{zcash} and an important part of the zk-SNARKS algorithm groth16\cite{groth2016size}, represents computations as a set of constraint conditions, namely linear equations and inequalities. Each equation has its own set of coefficients, while each variable represents an input or output value. These equations and inequalities describe the limiting conditions of the computation, implying that satisfying these conditions correctly calculates the corresponding output result for the given input sequence. R1CS includes a formal definition of constraint-based computation rules, which can be verified using a set of public parameters and a private input sequence. For a more detailed understanding of the formal definition of R1CS, refer to Vitalik's blog~\cite{buterin2016quadratic}.

\begin{definition} \label{def: r1cs_def}
    R1CS is a format for ZKP ACs.
    An R1CS is a conjunction of constraints, each of the form:
    \[(\vec{a} \cdot \vec{x}) \times (\vec{b} \cdot \vec{x}) = (\vec{b} \cdot \vec{x})\]
    where \(\vec{a},\vec{b}\) and \(\vec{c}\) are vectors of coefficients (elements of the prime field consisting of the integers modulo some prime), and \(vec{x}\) is a vector of distinct "pseudo-variables". Each pseudo-variable is either a variable, representing an element of the field, or the special symbol 1, representing the field element 1. \(\cdot\) represents taking the dot product of two vectors except that all additions and multiplications are done \(\pmod p\), and \(\times\) represents the product of 2 scalars \(\pmod p\). Using a pseudo-variabale of 1 allows a constant addend to be represented in a dot product.
\end{definition}

In Circom, each triplet of vectors in R1CS represents a mathematical constraint. These vectors consist of coefficients of variables found at corresponding positions in the solution vector $\vec{s}$. The solution vector $\vec{s}$ includes assignments for all variables present in the R1CS equations.

For example, a satisfied R1CS is shown in Fig.\ref{An R1CS example}:

\begin{figure}[!ht]
\centering
\includegraphics[width=.9\linewidth]{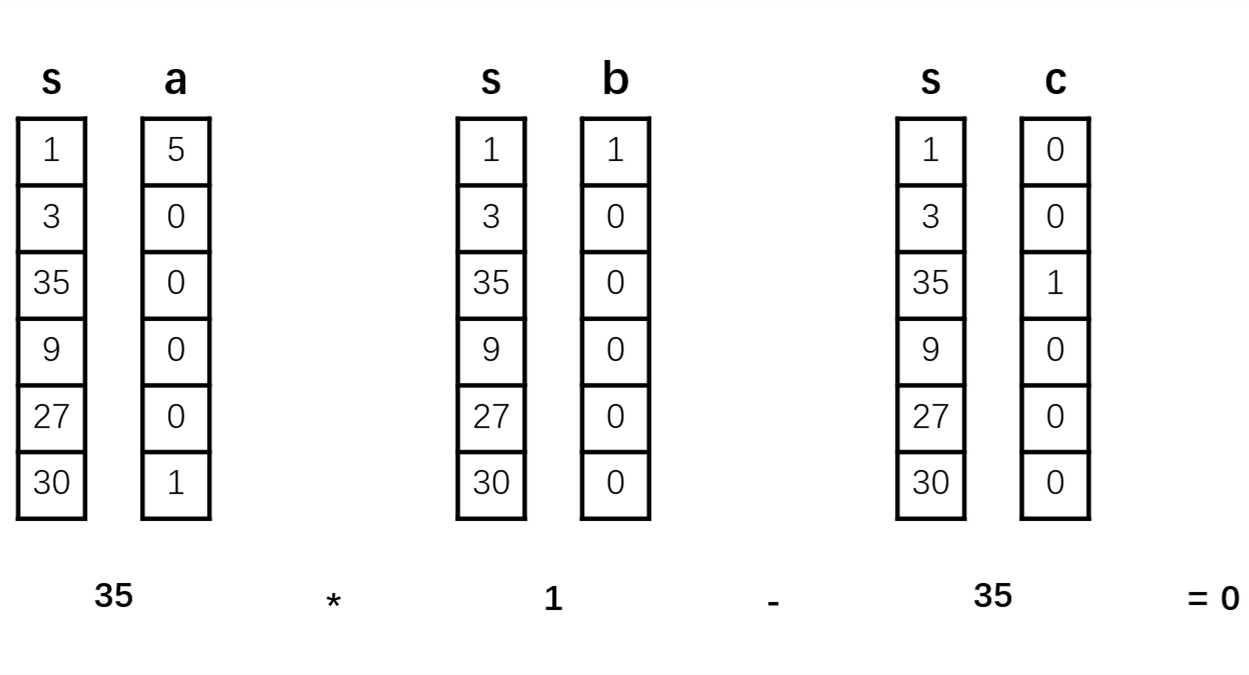}
\caption{A satisfied R1CS.} \label{An R1CS example}

\end{figure}

This constitutes a first-order constraint corresponding to a circuit multiplication gate. If we combine all constraints, we obtain a first-order constraint system.

R1CS is widely used in practical applications as a powerful computational model. It serves as an integral component of the groth16 algorithm, which is a popular version of zk-SNARK algorithms. R1CS plays a crucial role in improving developers' understanding of computer science and cryptography. Additionally, it offers crucial support for various privacy protection measures.54

In this paper, we propose the R1CS paradigm for constraint groups. It imposes constraints on the form and ordering of variable constraints.

\begin{definition} \label{def:r1cs_paradigm}
    R1CS paradigm is an R1CS satisfies the following requirements:
    \begin{enumerate}
        \item If a constraint in the R1CS paradigm contains multiplication between variables, it cannot have any other operators.
        \item  If a constraint in the R1CS paradigm does not contain multiplication between variables, it cannot contain intermediate variables generated by other linear constraints.
        \item The ordering of constraints and variables (defined in \textbf{Definition \ref{def: r1cs_def}}) in the R1CS paradigm must be consistent with the ordering method (\(\times\) is in front of \(+\)) in this paper.
    \end{enumerate}
\end{definition}

The specific adjustments required for converting a general constraint system into an R1CS paradigm will be explained through examples with concrete constraints. 

Requirement 1 suggests that complex quadratic constraints in the R1CS should be split into simpler forms.

For instance.
\begin{align*}
    a \times b+c+d=f \Longrightarrow \ a \times b=r, r+c+d=f\\
    5 \times a \times b=c \Longrightarrow 5\times a=r, r\times b=c
\end{align*}

Requirement 2 indicates that linear constraints in the R1CS system must be eliminated by removing intermediate variables defined by other linear constraints. For example,
\begin{align*}
    a+b=c, c+d=e \Longrightarrow a+b+d=e
\end{align*}

The specific sorting methods in requirement three will be discussed in later sections outlining the algorithm's steps.

\subsection{Data Flow Graph}
In a bipartite-directed graph, known as a data flow graph, there are two types of nodes: links and actors. Actors are utilized to represent various operations, while links serve as the means by which data is received by actors. Additionally, arcs allow links to transmit values to actors. The formal definition of this concept can be found in Dennis' paper \cite{dennis2005first}.

\begin{definition}  \label{def:data_flow_graph_def}
     A data flow graph is a bipartite labeled graph where the two types of nodes are called actors and links.
     \begin{equation}
         G=\langle A \cup L, E\rangle
     \end{equation}

     where

     $$
     \begin{aligned}
     &A={a_1, a_2, \ldots, a_n} &\text{is the set of actors} \\
     &L={l_1, l_2, \ldots, l_m} &\text{is the set of links} \\
     &E \subseteq (A \times L) \cup (L \times A) &\text{is the set of edges}.
     \end{aligned}
     $$
\end{definition}
A more detailed description can be found in \cite{treleaven1982data}.
\subsection{Weighted Pagerank Algorithm}
In this paper, we adopt the weighted PageRank algorithm to compute the weight of each node in the data flow graph~\cite{xing2004weighted}.

Pagerank algorithm is a method used for computing the ranking of web pages in search engine results. It was initially proposed by Larry Page and Sergey Brin, co-founders of Google, in 1998 and has since become one of the most essential algorithms in the field of search engines.\cite{page1998pagerank}

The algorithm assesses online web pages to determine their weight values, which it then utilizes to rank search results. PageRank is based on the notion that the weight of a web page is influenced by both the quantity and quality of the other web pages that link to it.

The main steps of the Pagerank algorithm are as follows:
\begin{enumerate}
\item Building the graph structure: First, the web pages and links on the internet must be converted into a graph structure. In this structure, each web page corresponds to a node and each link to a directed edge that points to the linked web page.

\item Computing the initial scores of each page: In Pagerank, the initial score of each page is set to 1. This means that initially, each node has an equal score.

\begin{figure}[!ht]

\centering
\includegraphics[width=0.8\linewidth]{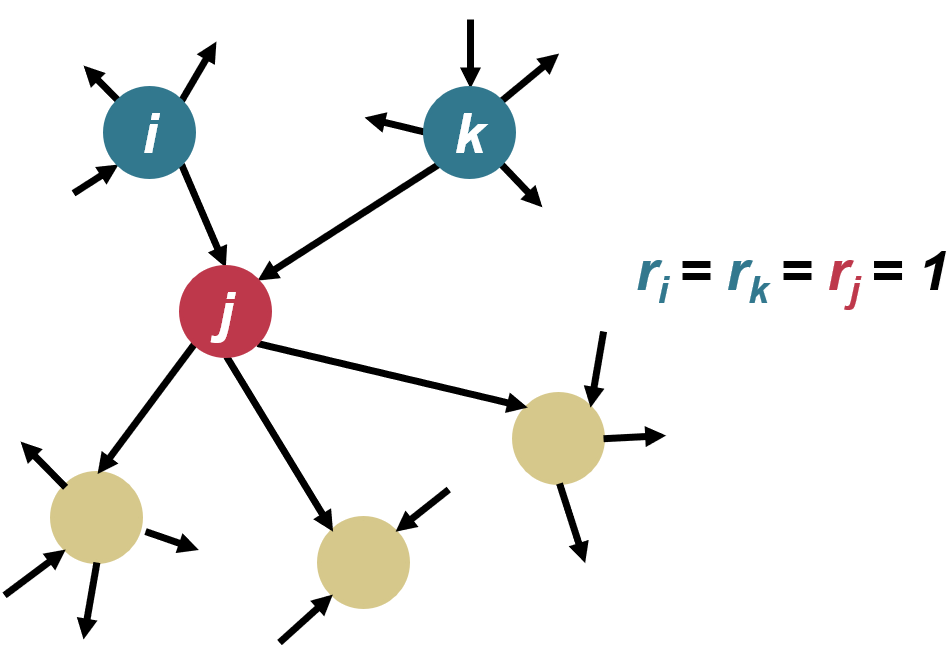}
\caption{The initial state of Pagerank algorithm.} \label{initial state}

\end{figure}
\item Iteratively computing the scores of each page: Each node's score is iteratively calculated based on its incoming links and averaged onto its outgoing links at each iteration.

\item Considering the number and quality of links: In addition to the relationships between nodes, Pagerank considers the number and quality of links pointing to a web page. Links from high-quality websites may carry more value than those from low-quality sites. Therefore, when computing scores, the algorithm weights links according to their number and quality.

\begin{figure}[!ht]
\centering
\includegraphics[width=0.8\linewidth]{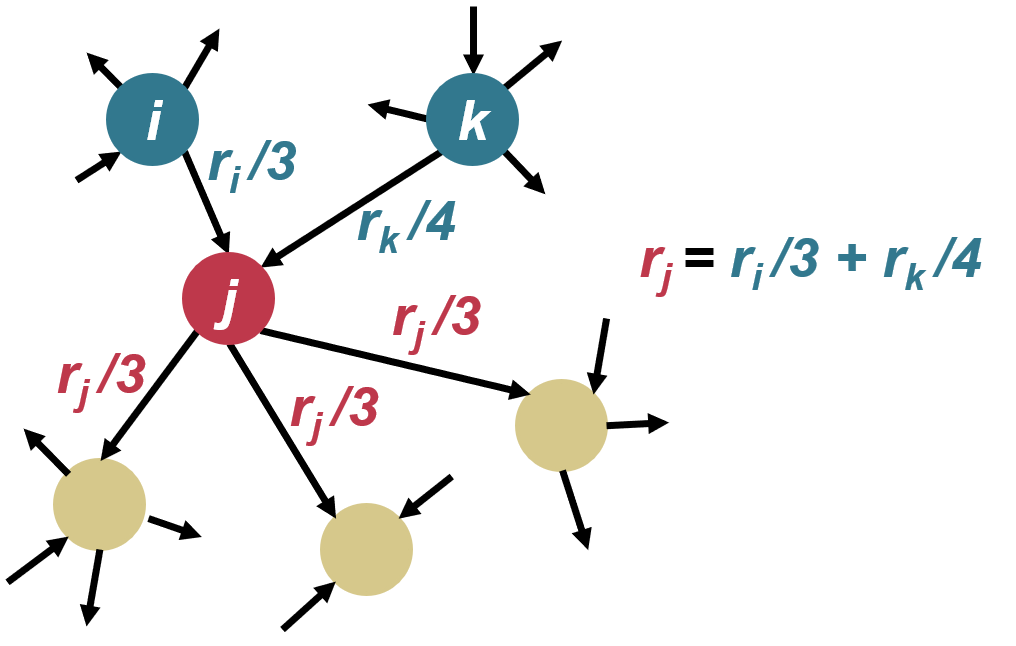}
\caption{The interation of Pagerank algorithm.} \label{interation state}

\end{figure}
\item Iterating until convergence: When the score of a node stabilizes, the algorithm stops iterating. This indicates that the final scores of all nodes have been determined and can be used to rank search results.
\end{enumerate}

The Weighted PageRank algorithm differs from the standard PageRank algorithm in that it incorporates the weight of each link as a factor, resulting in a more precise evaluation of a webpage's importance. Considering the importance of pages, the original PageRank formula is modified as

\begin{equation}
PR(u)=(1-d)+d \sum_{v \in B(u)} P R(v) W_{(v, u)}^{i n} W_{(v, u)}^{o u t}
\end{equation}

In this equation, $W_{(v, u)}^{i n}$ and $W_{(v, u)}^{o u t}$  are the weight of $link(v, u)$ calculated based on the number of inlinks and outlinks of page $u$ and the number of inlinks
of all reference pages of page $v$.

In this paper, we aim to use this algorithm to obtain more accurate weight values for each node in the data flow graph.

section{Overview}
In this section, we will introduce the procedure of normalization through the process of converting R1CS introduced in Vitalik's blog~\cite{buterin2016quadratic} in the algorithm: 

Constraint Set:
\begin{align*}
&A=\begin{pmatrix}
0 & 1 & 0 & 0 & 0 & 0\\
0 & 0 & 0 & 1 & 0 & 0\\
0 & 1 & 0 & 0 & 1 & 0\\
5 & 0 & 0 & 0 & 0 & 1
\end{pmatrix}
B=\begin{pmatrix}
0 & 1 & 0 & 0 & 0 & 0\\
0 & 1 & 0 & 0 & 0 & 0\\
1 & 0 & 0 & 0 & 0 & 0\\
1 & 0 & 0 & 0 & 0 & 0
\end{pmatrix} \\
&C=\begin{pmatrix}
0 & 0 & 0 & 1 & 0 & 0\\
0 & 0 & 0 & 0 & 1 & 0\\
0 & 0 & 0 & 0 & 0 & 1\\
0 & 0 & 1 & 0 & 0 & 0
\end{pmatrix}
\end{align*} \\

\begin{figure}
 \begin{center}   
  \begin{minipage}[t]{.8\linewidth}
    \centering
    \includegraphics[width=0.8\textwidth]{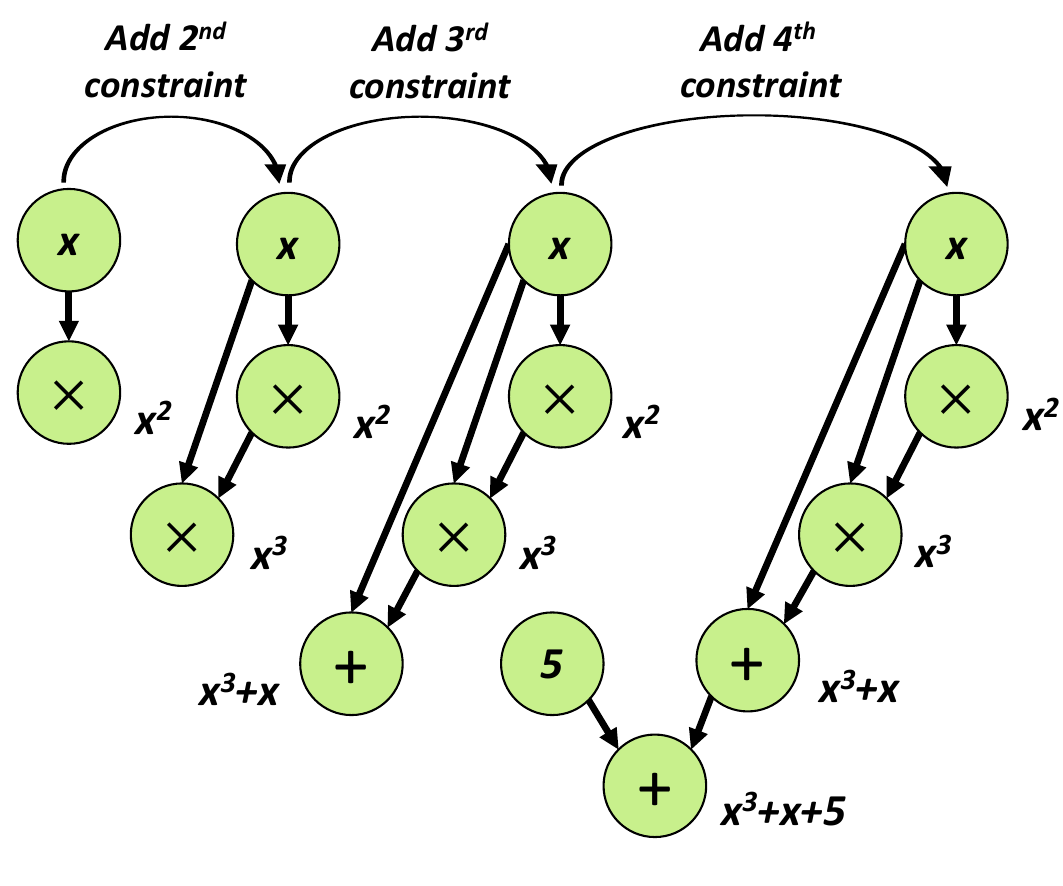}
    \caption{The procedure of constructing the data flow graph.}
    \label{fig:dfg_structure}
  \end{minipage}%
  \hspace{0.5cm}
  \begin{minipage}[t]{.8\linewidth}
    \centering
    \includegraphics[width=0.8\textwidth]{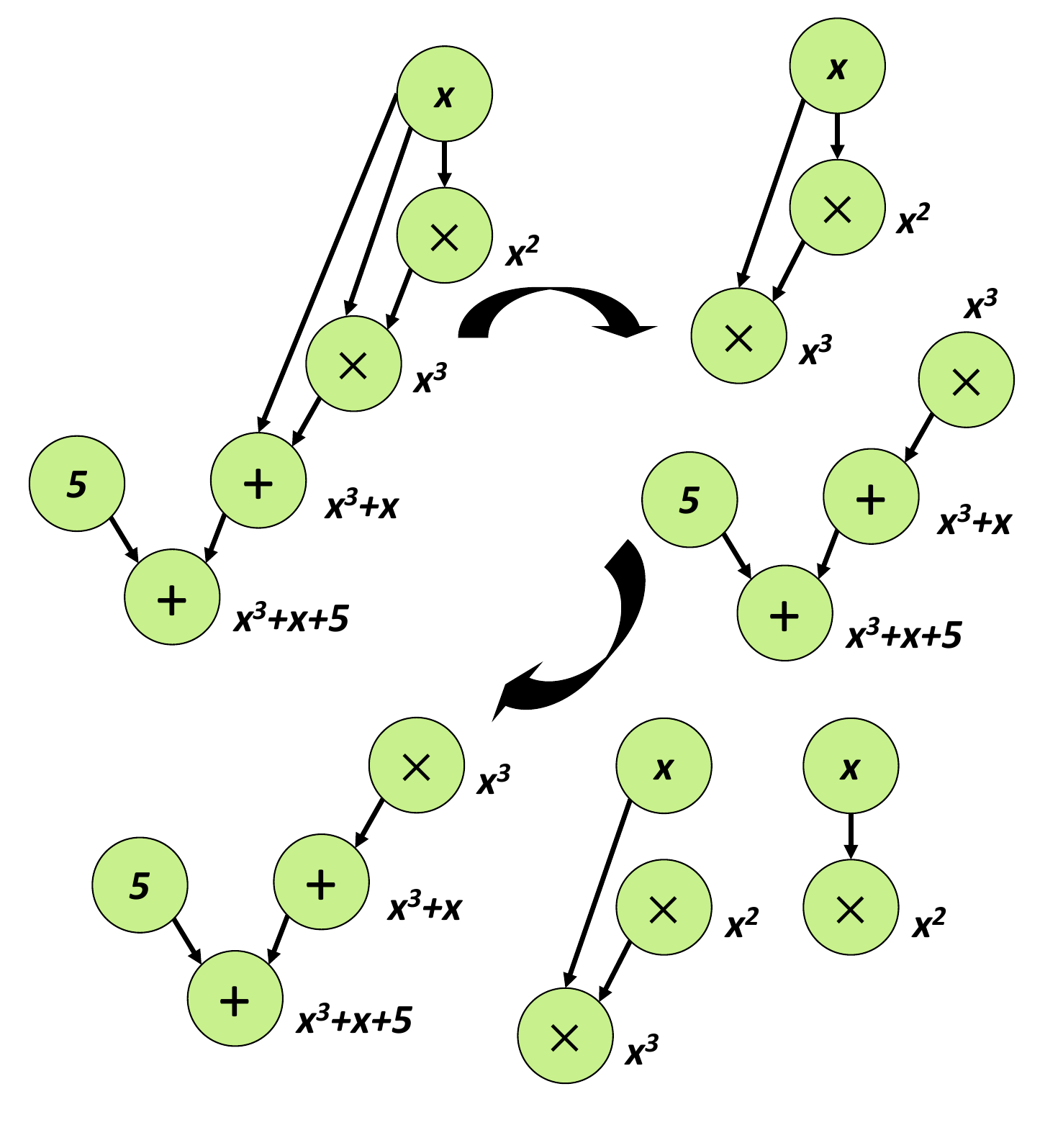}
    \caption{The procedure of tile selection.}
    \label{fig:tile_selection_procedure}
  \end{minipage}
  \end{center}
\end{figure}

Firstly, the arithmetic tree generation process involves creating an arithmetic tree for each constraint within the input R1CS constraint group, which is subsequently merged. The resulting arithmetic tree comprises common subformulas stored in a \emph{DAG}. The constructed data flow graph's structure is shown in figure \ref{fig:dfg_structure}, which illustrates how the arithmetic trees are combined to form the data flow graph.
                 
Subsequently, a tile selection algorithm is implemented based on the data flow graph, which divides the graph into tiles. The division of the entire graph into tiles is illustrated in figure \ref{fig:tile_selection_procedure}, depicting the overall procedure of the tile selection algorithm. The specifics of the tile selection process, including the form and selection logic, will be elaborated upon in subsequent chapters.

Next, the data flow graph is abstracted further with the selection of tiles as a reference. A new abstracted node in the data flow graph replaces linear constraints represented by tiles. The abstracted node can be represented as an affine mapping, which preserves the linear relationship between the variables, enabling faster computation of the intermediate values during the proof generation process. This abstraction procedure streamlines the proof generation process and reduces the computational cost of generating the proof.

\begin{figure}

\centering

\includegraphics[width=1\linewidth]{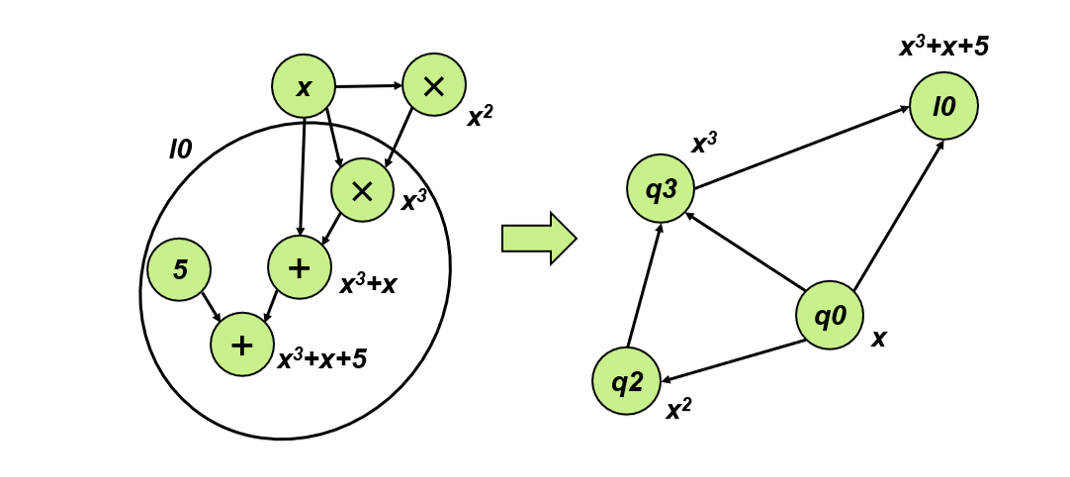}
\caption{The procedure of constructing the data flow graph.} \label{fig:abstraction_procedure}

\end{figure}

\begin{figure}

\centering

\includegraphics[width=0.5\textwidth]{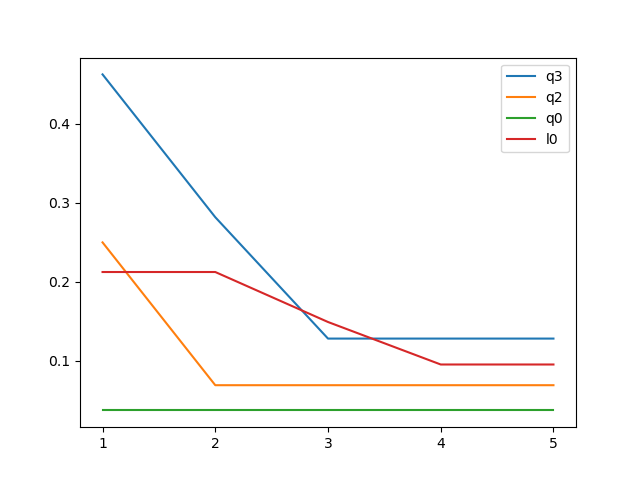}
\caption{The convergence process of each node.} \label{fig:w_pgrk}

\end{figure}

We calculate the weight of each node with coefficients of the constraint. Then we calculate the weights of the selected individual tiles using the improved Weighted PageRank algorithm. The convergence process of the PageRank values of four nodes in the abstract graph is depicted in figure \ref{fig:w_pgrk}.

Finally, constraints in the paradigm of R1CS are generated separately for each tile. And the constraints and variables are ranked by the node weights computed in the previous steps. 

Now we convert the input R1CS to its paradigm:

\begin{align*}
&A=\begin{pmatrix}
0 & 1 & 0 & 0 & 0 \\
0 & 1 & 0 & 0 & 0\\
5 & 1 & 0 & 1 & -1
\end{pmatrix}
B=\begin{pmatrix}
0 & 0 & 1 & 0 & 0\\
0 & 1 & 0 & 0 & 0\\
1 & 0 & 0 & 0 & 0
\end{pmatrix} \\
&C=\begin{pmatrix}
0 & 0 & 0 & 1 & 0\\
0 & 0 & 1 & 0 & 0\\
0 & 0 & 0 & 0 & 0
\end{pmatrix}
\end{align*}

\section{Normalization Algorithm}
In this section, we formally introduce various steps of the normalization generation algorithm and several data structures defined within the algorithm.

\subsection{Construction of RNode Graph}
Our study presents a novel data structure, RNode, that represents variable relationships within an R1CS arithmetic circuit. This structure, \emph{RNode}, facilitates efficient problem-solving by tracking interconnections among variables, which is formally defined as follows,

\begin{definition}\label{def:RNode_def}
An RNode is a node of two types in the data flow graph constructed in this normalization algorithm.

    \begin{equation}
    \begin{aligned}
        RNode &= ConstNode \cup VarNode\\
        ConstNode&=\{ConstValue, Operation, Father, Child\}\\
        VarNode&=\{Operation, Father, Child\}\\
    \end{aligned}
    \end{equation}

     where

    $$
    \begin{aligned}
     &\forall c\;\text{is a ConstNode} \cap c.Operation = Null, c.child = \emptyset  \\
     &\forall c\;\text{is a ConstNode} \cap c.Operation \in \{Add, Mul\}, c.father = \emptyset 
    \end{aligned}
    $$

\end{definition}

RNodes can be categorized into two types based on the variables they represent. The first category represents the original variables in the solution vector of the R1CS and the intermediate variables produced during the construction of the arithmetic circuit. The second category represents the constants in the data flow graph. Each RNode includes both an operator and a computed result, which store the calculation method between its two parent nodes and represent the calculated result of the subtree rooted in itself.

The generation of the RNode Graph involves 3 main stages:

\begin{enumerate}
\item Transform each constraint into an equation of $a*b=c$, as required by the R1CS constraints. 
\item Convert each constraint in the original R1CS constraint into an equation.
\item Organize the resulting equations containing common sub-expressions into a DAG-structured expression tree.
\end{enumerate}

The core logic of the RNode Graph generation algorithm is similar to the procedure we previously mentioned for constructing the RNode Graph.

The RNode is a data structure used to store information about the variables in an R1CS during the construction of the RNode Graph. Unlike typical nodes in an expression tree, each RNode stores information about both the variable and operator involved in a given operation, allowing each operator's output to be considered an intermediate variable and making it more closely aligned with the properties of an R1CS constraint set.

In practice, the merging and splitting of constraints poses a challenge in determining the equivalence of R1CS constraint sets. However, our RNode graph generation algorithm has observed that this merging or splitting does not lead to substantial differences. When constraints are merged, one variable is subtracted from the original constraint set. For instance, the merging process can be exemplified by the following example.

\begin{equation*}
\begin{pmatrix}
1 & 1 & 0 & 0\\
0 & 1 & 1 & 0
\end{pmatrix}
\cdot
\begin{pmatrix}
1 & 0 & 0 & 0\\
1 & 0 & 0 & 0
\end{pmatrix}
=
\begin{pmatrix}
0 & 0 & 1 & 0\\
0 & 0 & 0 & 1
\end{pmatrix}
\end{equation*}
\begin{equation*}
\downarrow
\end{equation*}
\begin{equation*}
\begin{pmatrix}
1 & 2 & 0 & 0
\end{pmatrix}
\cdot
\begin{pmatrix}
1 & 0 & 0 & 0
\end{pmatrix}
=
\begin{pmatrix}
0 & 0 & 0 & 1
\end{pmatrix}
\end{equation*}
\begin{equation*}
\downarrow
\end{equation*}
\begin{equation*}
\begin{pmatrix}
1 & 2 & 0
\end{pmatrix}
\cdot
\begin{pmatrix}
1 & 0 & 0
\end{pmatrix}
=
\begin{pmatrix}
0 & 0 & 1
\end{pmatrix}
\end{equation*}

However, the subtracted variable will be added back into the RNode Graph as an intermediate node in the sum-product expression during the construction of the RNode Graph. The reverse is also true.

Further abstraction of the RNode Graph is required to eliminate the difference observed in the graphs generated by equivalent R1CS due to the different variable ordering in the constraint set. The main difference is observed in the order of addition in constructing the expression of continuous addition. At this stage, we do not have sufficient information to determine the sequential execution order. During the algorithm execution flow, two variables are randomly selected and added together, resulting in a different structure in the graph.  

\subsection{Tile Selection}
Here, we categorize tiles into three types and give the formal definition of \textbf{tile}.
\begin{definition}\label{def:tile_def}
Tile is a tree-like subgraph of the RNode Graph, representing a constraint in R1CS.
\begin{enumerate}
    \item Quadratic: Tiles with the form $x * y = z$, where $x$, $y$, and $z$ are variables.
    \item MulLinear: Tiles whose root is obtained by multiplying its two parents, with at least one of the parents being a constant.
    \item AddLinear: Tiles whose root is obtained by adding its two parents.
\end{enumerate}
\end{definition}

Tile is essentially a set of linear equations generated by applying certain constraints to variables in the R1CS. We can use tiles as building blocks to construct a normalized R1CS that is both correct and scalable.
 
During tile selection, we divide the data flow graph from the previous step into 3 types. While AddLinear and MulLinear tiles are generated by linear constraints and are essentially linear tiles, their logical processing differs significantly. Hence, we discuss them as two separate types.
\begin{enumerate}
    \item We temporarily put aside the constraint merging step until we obtain more information about the tree in subsequent steps.
    \item If there is a need to generate merged formulas later, it can be achieved simply by applying a fixed algorithm to the unmerged formulas.
    \item The implementation of the tile selection algorithm is relatively simple.
\end{enumerate}

When the RNode graph is generated, we select a node with no successors and use it as the root to partition a subgraph from the RNode Graph as a tile. We then remove this tile from the RNode Graph and repeat this loop until the entire graph is partitioned. Detailed code will posted in github. 

The difference between data flow graphs generated by equivalent R1CS constraint systems lies in the order of node additions when processing linear tiles. However, the nodes added within a linear constraint, when considered as a set, are equivalent. This implies that the different addition order only affects the traversal order of the nodes. Therefore, if we consider the selected linear tiles as the products of the selected nodes and their respective coefficients, the chosen sets of linear tiles from equivalent R1CS constraint systems are the same. In other words, there is no distinction between the selected sets of tiles for equivalent R1CS constraint systems. Revised paragraph: The difference between data flow graphs generated by equivalent R1CS constraint systems lies in the order of node additions when processing linear tiles. However, the nodes added within a linear constraint, when considered as a set, are equivalent. This implies that the different addition order only affects the traversal order of the nodes. Therefore, the selected sets of linear tiles from equivalent R1CS constraint systems are the same if we consider the selected linear tiles as the products of the selected nodes and their respective coefficients. In other words, there is no distinction between the selected sets of tiles for equivalent R1CS constraint systems.

\subsection{Graph Abstraction}
We further abstract the data flow graph based on the selected tiles to eliminate the differences between various equivalent R1CS constraint systems. Specifically, we abstract linear tiles using a new node. By doing so, we mask the difference in addition order within linear tiles in the RNode Graph and transform the relationship between external nodes and specific nodes within the linear tile into a relationship between external nodes and the tile to which the particular node belongs. In this abstracted data flow graph, the types of edges are as follows:

\begin{enumerate}
    \item Non-linear tile abstract node to non-linear tile abstract node: The two vertices already existed in the original RNode graph. This edge type remains consistent with the original RNode graph.
    \item Non-linear tile abstract node to linear tile abstract node: This edge type exists only if there are non-abstract nodes in the linear tile represented by the abstract node.
    \item Linear tile abstract node to linear tile abstract node: This edge type exists only if the two abstract nodes represent linear tiles that share common non-abstract nodes.
\end{enumerate}

\subsection{Tile Weight Calculation}
In this step, we use the Weighted PageRank algorithm to calculate the scores of each vertex in the data flow graph. 

The previous steps eliminated the differences in the data flow graphs generated by equivalent R1CS through the abstraction of linear tiles. In the following step, constraints are generated on a tile-by-tile basis, and a criterion for tile order is proposed to sort the generated constraints. 

In the algorithm proposed in this paper, the Weighted PageRank algorithm is used to calculate the weights of each node in the abstracted data flow graph, which are then used as the basis for calculating the weights of the corresponding constraints for each tile. Compared to the traditional PageRank algorithm, this algorithm assigns weights to every edge in the graph and adjusts the iterative formula for node weights. In the Weighted PageRank algorithm, as mentioned in the former section, the formula for calculating node scores is defined as follows:

\begin{equation}
PR(u)=(1-d)+d \sum_{v \in B(u)} PR(v) W_{(v, u)}^{ in} W_{(v, u)}^{out}
\end{equation}

In this algorithm, the primary purpose of using the Weighted PageRank algorithm is to reduce the symmetry of the abstracted data flow graph. The structure of the graph is significantly simplified in the previous step through the simplification of linear tiles. However, some structures in the graph still contain symmetric nodes. If general algorithms are used to calculate the weights of nodes, these symmetric nodes may be assigned the same weight, which can lead to problems in subsequent constraint generation and sorting. To address this issue, the Weighted PageRank algorithm is employed to calculate weights for different nodes. This helps increase the asymmetry of the graph and minimize the occurrence of nodes with the same score.

In this algorithm, the iterative formula for scores in the Weighted PageRank algorithm is further adjusted. The node weights retained in the original data flow graph are set to 1, while for nodes abstracted from linear constraints, their weights are calculated through a series of steps.

First, for a linear constraint:

\begin{equation}
\sum^{n}_{i=1} a_ib_i=c
\end{equation}

Convert it to:
\begin{equation}
\sum^{n}_{i=1} a_ib_i-c=0
\end{equation}

Finally, the variance of the normalized coefficients is utilized as the weight of the abstract node representing the linear constraint.

\begin{equation}
W=\frac{\sum_{i=1}^{n}\left(a_{i}-\frac{\sum_{i=1}^{n} a_{i}-1}{n+1}\right)^{2}+\left(-1-\frac{\sum_{i=1}^{n} a_{i}-1}{n+1}\right)^{2}}{\left(\frac{\sum_{i=1}^{n} a_{i}-1}{n+1}\right)^{2}}
\end{equation}

In this algorithm, the iterative formula for the node score in the Weighted PageRank algorithm is given by:

\begin{equation}
PR(u)=(1-d)+d \sum_{v \in B(u)} PR(v) W^{u} W^{v}
\end{equation}

The scores for each node are calculated to determine the ranking of each constraint in the R1CS form.

\subsection{Constraint Generation}
In this step, we generate constraints in descending order of weight, with tiles as the unit, starting with quadratic constraints and followed by linear constraints.

\begin{figure}

\centering

\includegraphics[width=0.5\textwidth]{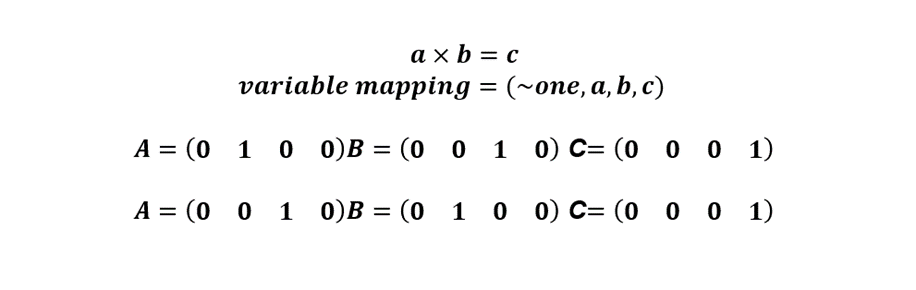}
\caption{Two equivalent expressions of a quadratic constraint.} \label{equivalent_quadratic}

\end{figure}

In this phase, a portion of the variable ordering was determined through the quadratic tile. In the tripartite matrix group of the R1CS paradigm, the three-row vectors corresponding to quadratic constraints each have only one non-zero coefficient. Still, in matrices A and B, the two-row vectors corresponding to the same constraint are equivalent in the constraint representation, consistent with the commutativity of multiplication. This results in two equivalent expressions for the same quadratic constraint. Figure \ref{equivalent_quadratic} illustrates an example of equivalent terms for a quadratic constraint.

In the abstract data flow graph, all vertices representing variables appearing in quadratic constraints are retained, making it possible to determine the choice of non-zero coefficients in the corresponding row vectors of matrices A and B in the tripartite matrix group of the R1CS paradigm based on the weights of each variable. The variables with higher weights are assigned to matrix A and given smaller indices in the variable mapping.

The sorting rules for the ordering of variables that appear in quadratic constraints can be summarized as follows:
\begin{enumerate}
\item Sort variables based on the highest weight value among all quadratic constraints in which they appear, such that variables with higher weight values have smaller indices in the variable mapping.

\item For variables that appear in the same constraint and have the same highest weight value sort them based on their scores in the Weighted PageRank algorithm for the nodes in the data flow graph corresponding to the variables. Variables with higher node scores will have smaller indices in the variable mapping and be assigned to the corresponding row vector in matrix A of the tripartite matrix group.
\end{enumerate}

\subsection{Adjustment of Linear Constraints}
At this stage, the partition and ordering of constraints within a constraint group have been established, and the ordering of variables that appeared in quadratic constraints has also been determined. However, it is necessary to adjust the order of newly introduced variables in linear tiles in this step. As the example below demonstrates, introducing multiple new variables within a linear tile can result in disorderly sorting. This is because, in the previous actions, the specific structures of linear tile constraints were abstracted to eliminate differences in the RNode Graph. 
\begin{align*}
&A=\begin{pmatrix}
0 & 1 & 0 & 0 & 0 & 0 \\
0 & 1 & 0 & 0 & 0 & 0\\
5 & 1 & 0 & 1 & -1& 1
\end{pmatrix}
B=\begin{pmatrix}
0 & 0 & 1 & 0 & 0 & 0\\
0 & 1 & 0 & 0 & 0 & 0\\
1 & 0 & 0 & 0 & 0 & 0
\end{pmatrix}\\
&C=\begin{pmatrix}
0 & 0 & 0 & 1 & 0 & 0\\
0 & 0 & 1 & 0 & 0 & 0\\
0 & 0 & 0 & 0 & 0 & 0
\end{pmatrix}
\end{align*}
\begin{align*}
&A=\begin{pmatrix}
0 & 1 & 0 & 0 & 0 & 0 \\
0 & 1 & 0 & 0 & 0 & 0\\
5 & 1 & 0 & 1 & 1& -1
\end{pmatrix}
B=\begin{pmatrix}
0 & 0 & 1 & 0 & 0 & 0\\
0 & 1 & 0 & 0 & 0 & 0\\
1 & 0 & 0 & 0 & 0 & 0
\end{pmatrix} \\
&C=\begin{pmatrix}
0 & 0 & 0 & 1 & 0 & 0\\
0 & 0 & 1 & 0 & 0 & 0\\
0 & 0 & 0 & 0 & 0 & 0
\end{pmatrix}
\end{align*}

Therefore, we introduce a new method to order the newly introduced variables in linear tiles. For each new variable introduced in a linear tile, its weight is calculated as

\begin{equation}
    weight = \sum_{other\ linear\ tiles} \left | field * weight\ of\ linear\ tile \right |
\end{equation}

The new variables are sorted based on their weights. If the weights are the same, the coefficients of the variable in its linear tile are considered for comparison. The appearance of new variables in other linear tiles, to some extent, reflects their importance in the entire constraint group. Additionally, suppose certain new variables only appear in their constraints. In that case, their weights will be zero, and their ordering will only affect the constraints generated by their linear tile without changing the ordering of other constraints. Therefore, they can be sorted in descending order based on their coefficients alone.

\section{Evaluation}
In this section, we introduced the self-designed benchmark used in this paper. We evaluated the effectiveness of the paradigm generation algorithm by analyzing the test results and the intermediate outputs.
\subsection{Benchmark Design}
To evaluate the proposed algorithm in this paper, we implemented the entire process of paradigm generation explained in the former section using Python to verify its results.

Due to the lack of related research, this field has no comprehensive benchmark. Therefore, we summarized some rules for generating equivalent R1CS constraint groups based on the logic the mainstream Circom compiler used to create R1CS and designed a more comprehensive benchmark based on these rules. The benchmark includes the following main categories depending on the reflected situation:

\begin{enumerate}
    \item Replacement of variable order in R1CS.
    \item Transformation of constraint order in R1CS.
    \item Introduction of multiple new variables in a single linear constraint in R1CS.
    \item Introduction of new variables in multiple linear constraints in R1CS, with shared new variables.
    \item Merging and splitting of constraints in R1CS.
\end{enumerate}

The different categories in the benchmark correspond to the other reasons for generating equivalent R1CS. Each category contains 2-3 basic R1CS constraints. To comprehensively test the robustness and correctness of the algorithm, 5-6 equivalent R1CS constraint groups are generated for each R1CS based on the respective reasons. The equivalent constraint groups of each constraint group are paired and inputted into the algorithm to verify whether the algorithm can generate consistent and R1CS-compliant output results defined in definition \ref{def:r1cs_paradigm}, when processing different equivalent constraint groups.

The publicly available data set used in this study can be found at the following GitHub repository: \url{https://github.com/Ash1sc/R1CS\_normalization\_benchmark}. This repository contains the raw data that was utilized for testing purposes. It is important to note that the data set is licensed under the GNU General Public License version 3.0 (GPL-3.0). This license allows for the data set and scripts to be freely distributed, modified, and used, with the condition that any derived works are also licensed under the GPL-3.0 and that the original copyright and license information is retained. If you would like to learn more about the details of the GPL-3.0, please visit \url{https://www.gnu.org/licenses/gpl-3.0.en.html}.
\subsection{Result Evaluation}
Table \ref{tab:res} shows the result of the experiments.

\begin{table}[!ht]
\centering
\caption{Experimental Results of Equivalent R1CS Constraint Group Conversion}
\label{tab:res}
\begin{tabular}{p{2.5cm}p{1.5cm}<{\centering}p{1.25cm}<{\centering}c}
\toprule
Reasons for Generating Equivalent R1CS Constraints & Number of Groups & Successfully Generated Groups & Pass Rate \\ \midrule
Replacement of variable order in R1CS. & 55 & 55 & 100\% \\ \\
Reordering of constraint sequences in R1CS. & 21 & 21 & 100\% \\ \\
Introduction of multiple new variables in a single linear constraint in R1CS. & 15 & 15 & 100\% \\ \\
Introduction of multiple new variables with shared usage in multiple linear constraints in R1CS. & 15 & 15 & 100\% \\ \\
Merging and splitting of constraints in R1CS. & 6 & 6 & 100\% \\ \bottomrule
\end{tabular}
\end{table}

Observation shows that the generated paradigms meet the requirements of the R1CS paradigm mentioned above and have equal semantics as the original R1CS constraint groups.

Through analysis of the intermediate outputs at each stage of the conversion process, it was found that for equivalent R1CS constraint groups generated by reordering constraints, the only difference in the resulting data flow graphs lies in the order in which nodes representing intermediate variables are created. This is due to different processing orders of each constraint during traversal, leading to other orders of introducing intermediate variables in each constraint.

For equivalent R1CS constraint groups generated by variable replacement, the differences lie in the order in which RNodes representing initial variables in the R1CS are developed and in the order of addition in the summation chain structure caused by differences in variable order in linear constraints. However, these changes do not affect the selected tile set, and the same data flow graph is obtained after abstraction.

In the final step of the algorithm, we proposed a novel variable ordering method to solve the sorting confusion issue when multiple variables are introduced to a linear constraint.Experimental results demonstrate that our algorithm is capable of correctly identifying these variables and produces a variable mapping sequence that conforms to the definition.

After the abstraction of the data flow graph, the splitting and merging of linear constraints can lead to changes in the order of addition in the summation chain. However, this issue is resolved. When equivalent R1CS constraint groups are created through the merging and splitting of constraints, the only discrepancy in the resulting data flow graphs is with regards to the vertices representing intermediate variables. These vertices can either represent existing variables or intermediate variables introduced during the data flow graph's creation. Nevertheless, the structure of the data flow graph remains unaffected.

\section{Conclusion}
In this paper, we propose an algorithm based on data flow analysis to construct a paradigm for R1CS. The correctness and equivalence of R1CS have long been challenging to study due to the diversity and flexibility of constraint construction methods. Our algorithm aims to eliminate the differences between equivalent R1CS constraint systems through a series of abstraction processes. We introduce ordering methods for internal variables and constraints in R1CS, which provide reference information for the final paradigm generation. Through experimental results, we demonstrate that our algorithm can identify equivalent R1CS resulting from constraint merging, intermediate variable selection, and variable and constraint reordering, and transform them into a unique paradigm. Therefore, R1CS plays an indispensable part in zk-SNARKS.
Further work includes establishing rules for merging constraints and a more comprehensive benchmark. This requires us to conduct more in-depth research and exploration of the generation rules of R1CS to improve the algorithm.
\printbibliography
\end{document}